\documentclass[aps,prl,reprint,floatfix, footinbib, amsmath,amssymb,superscriptaddress]{revtex4-2}

\usepackage{graphicx}
\usepackage{epstopdf}
\usepackage{color}
\usepackage{hyperref}
\usepackage{bm}
\usepackage{printlen}
\usepackage{units}

\definecolor{cornflowerblue}{RGB}{100, 149, 237}
\usepackage[table]{xcolor}

\renewcommand{\vec}[1]{\mathbf{#1}}

\hypersetup{hidelinks}

\begin{document}

\title{Topological control of quantum speed limits}

\author{Alexander  Kruchkov}  

\affiliation{Department of Physics, Princeton University, Princeton NJ 08544, USA}

\thanks{Email: alexkruchkov@princeton.edu}

\date{\today}

\begin{abstract}
Quantum Fisher Information (QFI) is a measure quantifying the \textit{sensitivity} of a quantum state with respect to changes in tuning parameters in quantum metrology, and defining quantum speed limits. 
We show that even if the quantum state is completely dispersionless, QFI in this state remains momentum-resolved. 
We compute the QFI for  topological phases at integer filling and  demonstrate that each momentum-resolved term is fundamentally bounded by quantum geometric and topological invariants, with QFI maximum position being controlled by topological invariants. We  find bounds on quantum speed limit which scales as $\sqrt{|C|}$ in a dispersionless topological phase. We conclude that quantum platforms of high Chern numbers $|C| \gg 1$, such as those featuring twisted multilayered van der Waals heterostructures, significantly enhance capacity for quantum Fisher information, and provide practical control over quantum speed limits. 
\end{abstract}

\maketitle

\textit{Quantum Fisher information} is a measure quantifying the sensitivity of a quantum state with respect to changes in a parameter in quantum metrology \cite{Fisher1925, Helstrom1967, Helstrom1969, Holevo2011, Braunstein1994, Braunstein1996}.
For a quantum system perturbed by a time-dependent external electrostatic potential $ V_{\vec{q}}^{\text{ext}} $, the \textit{Bures distance} \cite{Bures1969, Uhlmann1976}  between the initial density matrix  $ \rho(0) $  and \( \rho(dt) \) is defined via the quantum Fisher information $f_Q(\vec{q})$ \cite{Braunstein1994, Balut2025},
\begin{align}
  ds^2 = \frac{1}{4} \sum_{\vec{q}} f_Q(\vec{q})\, |V_{\vec{q}}^{\text{ext}}|^2\, dt^2.
  \label{Bures}
\end{align}
Once a theoretical construct,  QFI has become a practical and experimentally measurable quantity \cite{Strobel2014, Hauke2016,exp1,exp2,exp3,exp4,exp5}.  
On the practical side, quantum Fisher information is related to \textit{quantum speed limits}, the fundamental bounds on the evolution rate of quantum states\cite{Toth2014, Pires2016, Frowis2012}, which are essential for estimating the ultimate speeds of quantum computation \cite{Margolus1998, Lloyd2000}.  The quantum Cramér–Rao bound (CRB) defines the fundamental precision limit for single-parameter estimation, tying the best possible measurement accuracy to the inverse of the quantum Fisher information (QFI)  \cite{Braunstein1994,Helstrom1967}. 
QFI also determines the critical driving rate for quantum Zeno effect \cite{Smerzi2012, Toth2014}: 
While it is usually impossible to probe a quantum system without disturbing it, the Quantum Zeno Effect (QZE) occurs upon  frequent measurements which suppress the evolution of a quantum system, preventing transitions between states; this happens when the measurement rate exceeds $\sim \sqrt{f_Q}$ \cite{Smerzi2012, Schafer2014}.  
There is a multifaceted interest in enhancing  capacity for quantum Fisher information.

On the other hand, quantum geometry is a powerful framework
 offering fresh insights into the fundamental properties of quantum systems \cite{Torma2023, Torma2022}. 
Many quantum phenomena, from unconventional superconductivity in flat topological bands to optical, thermal, and thermoelectric properties, are being reexamined through the lens of quantum geometry \cite{Torma2022, Ahn2022, Kruchkov2023, Tai2023}. 
More recently, entanglement measures and bounds on bipartite entanglement have been examined from quantum-geometric perspective \cite{ KRYU2024a, Tam2024}. In the core of this formalism lies the concept of quantum-metric tensor $\mathfrak G_{ij} (\vec k)$, which encapsulates both topological and geometric properties of quantum states, and defines the quantum distance between two adjacent quantum states \cite{Provost1980, Matsuura2010}. 
Recent research demonstrates that the quantum-geometric approach uncovers unexpected and often counter-intuitive effects, offering deeper insights into the complexity of topological quantum phases \cite{Yu2024}.
 Meanwhile, a new class of quantum materials has emerged by stacking and twisting one or more monolayers of van der Waals materials, enabling precise control over bandwidth, band gaps, topological and quantum-geometric properties \cite{Nuckolls2024}.

In this paper, we show that capacitance in quantum Fisher information in topological insulators is directly tied to engineering the underlying quantum geometry, with a lower bound on QFI defined by the associated topological invariants.
In particular, we establish that the leading terms take the form  
\begin{align}
f_Q (\vec q)  = f_Q^{[\text i]} + f_Q^{[\text{ii}]}+ ... 
\\
= A  q^2 - B q^4 + ... 	
\end{align}
where the coefficients $A $ and $B$ are each constrained by the Chern number.
This result follows directly from our main finding  which provides a general formula for the quantum Fisher information at integer filling of a  topological system with interactions. We discuss application to practical platforms, such as twisted transition metal dichalcogenides with flat topological bands, and derive the key formulas applicable to flat band models.  Finally, we discuss strategies for enhancing the quantum Fisher information capacity by involving higher-Chern van der Waals multilayers.  While traditionally QFI is treated for non-interacting systems at zero temperature, our approach allows for (i) extending the quantum Fisher information for interacting topological insulators and finite temperatures; (ii) naturally addressing the measurement of quantum Fisher information through the widespread techniques for measuring current-current correlators.

 { \textbf{ Formalism. } } Building on our previous work \cite{KRYU2023b,Kruchkov2023}, we extend the frequency-dependent Kubo formalism in topological phases  to include momentum-resolved responses to address the QFI \eqref{QFI-sigma}.      The main object for the present analysis is the momentum-resolved current-current correlator in Matsubara representation, 
\begin{align}
	\langle  \vec J (i \omega_n, \vec q) \vec J (i \omega_n, - \vec q) \rangle 
\end{align}
To achieve this, it is important to include quantum-geometric form-factors, as functions of $\vec q$, correctly in the calculation. For this, one needs to use current operators explicitly satisfying the charge conservation equation
$\partial_t \rho (\vec r) + \nabla \cdot \vec J(\vec r) =0 $. 
In the second-quantized form, the charge-conserving moment-resolved current operators take the following representation in the band basis \cite{Mckay2024} (in what follows below, we set $\hbar$$=$$1$): 
\begin{align}
\vec J(\vec{q}) = e \sum_{\vec{k}} \sum_{n,m} \int_{0}^{1} d\alpha \,
\langle 
u_{n \vec{k}} 
| \boldsymbol{ \mathcal  V}_{\vec{k} + (1 - \alpha) \vec{q}} | 
u_{m \vec{k} + \vec{q}} 
\rangle 
c^{\dagger}_{n \vec{k}} c^{}_{m \vec{k} + \vec{q}}. 
\label{conserved0}
\end{align}
Integrating over continuous dimensionless parameter $\alpha$ which parametrizes all possible momentum transfers ensures gauge invariance. 
We shall use expression \eqref{conserved0} in a symmetrized form  
\begin{align}
J_i(\vec{q}) = e \sum_{\vec{k}} \sum_{n,m}  \int \limits_{0}^{1} d \alpha 
\langle 
u_{n , \vec{k} -  \overline \alpha \vec{q}} 
|\mathcal  V^{i}_{\vec{k} } | 
u_{m \vec{k} + \alpha \vec q } 
\rangle  \nonumber
\\
\times 
c^{\dagger}_{n \vec{k} - \overline \alpha \vec q } c^{}_{m, \vec{k} + \alpha \vec q }, 
\label{conserved}
\end{align}
\noindent 
where we define $\overline \alpha \equiv 1 - \alpha  $ for notational brevity. 
 In the limit $\vec{q} \to 0$, this expression simplifies to the standard formula for the uniform current operator found in textbooks \cite{Mahan}.

Fundamentally, quantum Fisher information of an interacting system can be expressed in terms of quantum correlators, making it amenable to analysis via the Kubo formalism and measurable via a direct experimental probe. At zero temperature, the QFI can be elegantly expressed as \cite{Hauke2016,Shitara2016,Balut2025} 
\begin{align}
f_Q (\vec q) 	 = -  \frac{4 }{\pi} \int \limits_{0}^{\infty} d \omega  \, \text{Im} \chi  (\vec q, \omega) \tanh{\frac{\beta \omega }{ 2} } ,
\label{QFI-correlator}
\end{align}
Here $\beta = 1/T$ is inverse temperature; note that factor $4$ is introduced for convenience of Eq. \eqref{Bures}. 
  We note that the expression \eqref{QFI-correlator}, although accessible via linear response theory, differs from the static structure factor $S(\vec q)$ as an \textit{observable}   \cite{Nozieres1999}   
\begin{align}
 S(\vec q) 	 =  - \frac{1}{\pi} \int \limits_{0}^{\infty} d \omega  \, \text{Im} \chi  (\vec q, \omega) \coth{\frac{\beta \omega }{ 2} } ,  
\label{static-factor}
\end{align}
In essence, Eq.\eqref{static-factor} is  the fluctuation-dissipation theorem \cite{Nozieres1999}, while Eq.\eqref{QFI-correlator} is not. 
At low frequencies $\omega$, the relation $\coth\left(\frac{\beta \omega}{2}\right) \propto \frac{1}{\omega}$ implies that the integrand in \eqref{static-factor} is dominated by low-frequency dissipation on the right-hand side, corresponding to thermal fluctuations $S (\vec q)$ on the left-hand side. 
In contrast, integrand in Eq. \eqref{QFI-correlator}, with $\tanh{\frac{\beta \omega }{ 2} } \propto \omega$ at small $\omega$,  suppresses low-frequency dissipation, and prioritizes quantum-coherent features at higher frequencies, thus capturing   the system’s short-time response to external perturbations. Hence $4 S (\vec q) \ne  f_Q (\vec q)$. In fact, $4 S (\vec q) >  f_Q (\vec q)$.

 To calculate QFI in a topological system, one needs to evaluate the imaginary part of dynamical susceptibility $\chi  (\vec q, \omega)$.  A consequence of charge conservation $ \langle \vec q \cdot \vec J (\vec q, \omega) -  \omega  \rho (\vec q, \omega ) \rangle = 0 $ 
 is connection between dynamical susceptibility and longitudinal conductivity,  $q_{i} \sigma_{ij}(\vec q, \omega)q_j  = - i \omega e^2 \chi (\vec q, \omega) $ ($i, j= x,y$ through the text).  
 The generic expression \eqref{QFI-correlator} can be further simplified to Nozi{\'e}res form \cite{Nozieres1999}  assuming symmetries satisfying
 $\sigma_{xx} = \sigma_{yy}$, 
 $
 \chi (\vec q, \omega) =  \frac{i  q^2}{e^2} 	\frac{\sigma(\vec q, \omega) }{\omega}
$
where $\sigma(\vec q, \omega) = \sigma_{xx}(\vec q, \omega)$, $q=q_x$ are assumed (we use $q_x$ below for clarity). This yields a relation for the imaginary part $
\text{Im}	 \chi (\vec q, \omega) = - \frac{ \vec q^2}{e^2} 	\frac{\text{Re} \sigma(\vec q, \omega) }{\omega} $. 
One therefore arrives to a practical expression for QFI of an (interacting) system, 
\begin{align}
f_Q (\vec q) 	 = \frac{4 q^2}{\pi e^2}  \text{Re} \int \limits_{0}^{\infty}   \frac{  d \omega  \,  \sigma  (\vec q, \omega) \tanh{\frac{\beta \omega }{ 2} }}{\omega} .
\label{QFI-sigma}
\end{align} 
 Following this prescription, we consider response in  $\vec q = (q_x, 0)$.  
This generally yields QFI \eqref{QFI-sigma}  $q$-expansion of the form   
$
f_Q (q_x)   = A q^2_x + B q^4_x + ... 	
$
 which we derive below term by term.  Similarly, treating $yy$ response instead of $xx$ response in definition \eqref{QFI-sigma}  gives  $f_Q (q_y)   = A q^2_y + B q^4_y + ... 	$.  We consider below a response averaged over directions, so that $A$ and $B$ characteristics are same direction-independent, and the final answer takes form $
f  (q)  = A q^2 + B q^4 + \mathcal O (q^6) 	
$. We will use $q=q_x$ inter-exchangeably, unless specified otherwise.

 {\textbf{Quantum Fisher information in linear response theory}. In what follows below, we focus of the lattice systems resulting in band-resolved topological phases. We chiefly focus on the flat-band systems (such as twisted transition metal dichalcogenides), for which we use quantum transport formalism in the topological phase \cite{Kruchkov2023, KRYU2023a, KRYU2023b}, and extend it to wavelength-resolved response. We consider the  polarization diagrams, \begin{align}
\Pi_{xx} (\tau , \vec q) = \frac{e^2}{2} \sum_{\vec k} \sum_{nm}  & 
\int \limits_{0}^{1 }  d \alpha \,
G_{m, \vec k + \alpha \vec q } (\tau)  G_{n, \vec k -  \overline \alpha \vec q} (- \tau)
\nonumber 
\\
& \times \mathcal Q^{(\alpha)}_{nm} (\vec k, \vec q)  .
\label{polarization}
\end{align}
Here $G_{n\vec k}$ are Green's functions associated with $n \vec k $ Bloch state. 
We have introduced  tensor 
\begin{align}
\mathcal Q^{(\alpha)}_{nm} (\vec k, \vec q) 
  =  
\langle 
u_{n \vec{k} -  \overline \alpha \vec q} 
|\mathcal  V^{x}_{\vec{k} } | 
u_{m \vec{k} + \alpha \vec q} 
\rangle 
\langle 
u_{m \vec{k}  + \alpha \vec q} 
|\mathcal  V^{x}_{\vec{k} } | 
u_{n \vec{k}  - \overline \alpha \vec q} 
\rangle 
\label{geoqq}
\end{align}
with nontrivial $\vec q$-dependence which we address below. 

We further proceed to Matsubara transform of Eq. \eqref{polarization} to imaginary external frequencies $i \omega_0 $, we have
 \begin{align}
\Pi_{xx} (i \omega_0 , \vec q) = e^2  \sum_{\vec k} \sum_{nm} \sum_{i \omega_n }  \int \limits_{0}^{1 }  d \alpha 
G_{m, \vec k + \alpha \vec q} (i \omega_n )  
\nonumber
\\
\times 
 G_{n,\vec k -  \overline \alpha \vec q} (i \omega_n  + i \omega_0)  \mathcal  Q^{(\alpha)}_{nm} (\vec k, \vec q) . 
 \label{Pi_xx}
\end{align} 
For the details of the calculation we refer to Refs. \cite{Kruchkov2023,KRYU2023b}. 
The conductivity is then obtained through analytical continuation to the real frequency
\begin{align}
\sigma (\omega, \vec q ) = \frac{ \Pi_{xx} (\omega, \vec q)  }{i \omega} ,
\end{align} 
and taking into account that for the filled band $\omega =0$ diamagnetic term cancels with $\Pi (0, \vec q)$.

We consider $T =0$ ($\beta \to \infty$) asymptote  of Eq. \eqref{QFI-sigma},
\begin{align}
 f_Q (\vec q) 	
\underset{\beta \to \infty}{\simeq}  \frac{4 q^2}{\pi e^2}  \text{Re} \int \limits_{0^+}^{\infty}   \frac{  d \omega  \,  \sigma  ( \omega, \vec q)}{\omega}  .
\label{QFI-asym}
\end{align} 
Such approximation is possible as $ \tanh{\frac{\beta \omega }{ 2} }$ at $\beta \to \infty$ and $\omega>0$ behaves as a smoothened step function $\tilde \theta (\omega>1) =1$ and $\tilde \theta (\omega=0) =0$. However, one should use caution when considering  asymptote \eqref{QFI-asym}. A sufficient condition for $\beta \to \infty$ approximation of  $\int _{0^+}^{\infty} d \omega F(\omega) \tanh{\frac{\beta \omega }{ 2} }$ with  $\int _{0^+}^{\infty} d \omega F(\omega) $ is absolute integrability $\int _{0^+}^{\infty} d \omega |F(\omega) | < \infty $  and absence of strong singularities at $\omega =0$. We note that these conditions are fulfilled by virtue of generalized sum rules  for conductivities in topological insulators $\int _{0}^{\infty}   \frac{  d \omega  \,  \text{Re} \sigma  (\vec q, \omega)}{\omega^n} $ \cite{KRYU2023b}.  Physically, this approximation is justified by the suppression of transport below the gap  $\Delta_0$ characteristic of insulating phases. Accordingly, the zero-temperature asymptotic regime in Eq. \eqref{QFI-asym} should be understood as $\beta \Delta_0 \ll 1$. The main contribution to Eq. $\eqref{QFI-asym}$ comes from poles in complex plain with $|\omega_* | \approx \Delta$ (including but not limited to the lowest gap $\Delta_0$). Thus, QFI \eqref{QFI-asym}
encodes a \textit{sum rule}, capturing the geometric and topological information of the system \cite{KRYU2023b}. 
% \begin{align}
%f_Q (\vec q) 	 =  2 q^2  \sum_{\vec k} \sum_{nm}  d \alpha  \int \limits_{0}^{\infty} \mathcal Q^{\alpha}_{nm} (\vec k, \vec q)   \int \limits_{0}^{1 }    d \omega_1 \rho_{m, \vec k +\alpha \vec q } (\omega_1) 
%\nonumber
%\\
%\times  \int \limits_{0}^{\infty}  d \omega_2  \, \rho_{n, \vec k - \overline \alpha \vec q} (\omega_2 )  
% \frac{f (\omega_1) - f (\omega_2)}{\omega_1 - \omega_2}   .
% \label{key-result} 
%\end{align} 
%We thus derive one of the main results of this paper, and expression for quantum Fisher information in a topological phase. \textcolor{cornflowerblue}{Here $\rho_{n \vec k} (\omega) = \frac{1}{\pi} \text{Im}  G^{\text{R} }_{n \vec k} (\omega) $ is the (interacting) spectral weight of the Bloch state $n \vec k$,  defined by causal (retarded) Green's function $G^{\text R}_{n \vec k} (\omega) $, and $f (\omega)$ is Fermi-Dirac function. }
In the flat bands, we neglect dispersive contributions by considering  $\varepsilon (\vec k + \vec q)  \simeq \varepsilon (\vec k ) $, which therefore leads to 
 \begin{align}
f_Q (\vec q) 	 &  \simeq   4 q^2  \int \limits_{0}^{1} d \alpha \sum_{ \substack { n  \in \text{filled} \\  m \ne n } }   \sum_{\vec k}  \frac{\mathcal Q^{(\alpha)}_{nm} (\vec k, \vec q) }{\Delta^2_{mn} (\vec k) } .
\label{key-result}
\end{align}
  We further proceed to evaluation of the quantum Fisher information in showcase examples.

{ \textbf{Quantum Fisher information in the leading order for a non-interacting system.} }   We now consider the leading-in-$q$ order term  of $T=0$ QFI formula. For illustrative purpose, we apply it  to a dispersionless  topological system with no interactions. In tensors, we aim to keep spatial coordinates $i,j=x,y$ in superscripts, and band indices $n,m$ in subscripts, for bookkeeping.  
In the $\vec q =0$ limit, tensor \eqref{geoqq}
can be expressed through the multi-band quantum-geometric tensor $\mathfrak G^{ij}_{nm} (\vec k) $ \cite{Matsuura2010, Kruchkov2023} 
\begin{align} 
\mathcal Q^{(\alpha)}_{nm}( \vec k, \vec q =0 ) & =  \langle 
u_{n \vec{k}} 
|\mathcal  V^x_{\vec{k}} | 
u_{m \vec{k} } 
\rangle 
 \langle 
u_{m \vec{k} } 
|\mathcal  V^x_{\vec{k} } | 
u_{n \vec{k} } 
\rangle 
\nonumber 
\\
& \equiv   \Delta_{nm}^2 (\vec k)  \mathfrak G^{xx}_{nm} (\vec k) 
\label{Qmn},
\end{align}
where $\Delta_{mn} (\vec k) = \varepsilon_m (\vec k) - \varepsilon_n ({\vec k} )$, with $\varepsilon_m (\vec k) \equiv \varepsilon_{m \vec k}$ is the energy dispersion of the $m$th Bloch state, and multiband quantum-geometric tensor
 \begin{align}
\mathfrak G^{ij}_{nm} = 
\langle 
\partial_i u_{n \vec k} | u_{m \vec k} \rangle 
\langle 
u_{m \vec k} | \partial_j u_{n \vec k}
\rangle, 
 \end{align}
 which is related to the single-band quantum-geometric tensor  $\mathfrak G^{ij}_{n}$ through identity
  \begin{align}
\sum_{m\ne n} \mathfrak G^{ij}_{nm} =  \mathfrak G^{ij}_{n}   = \langle 
\partial_i u_{n \vec k} | 1 - \mathcal P_{n \vec k} | \partial_j u_{n \vec k}
\rangle,  
 \end{align}
  with $\mathcal P_{n \vec k} = | u_{n \vec k} \rangle  
\langle 
u_{n \vec k}|$.  Taking the real part of this expression,  one arrives to the quantum metric (Fubini-Study metric) of the $n$th Bloch band $\mathcal G^{xx,yy}_n (\vec k)$, while the imaginary part is related to the Berry curvature $\mathcal F^{xy}_n (\vec k)$,
\begin{align}
 \mathcal G^{ij}_n (\vec k) = \text{Re} \mathfrak G^{ij}_n (\vec k), 
\ \ \ \ \ 
\mathcal F^{xy}_n (\vec k) = -2 \text{Im} \mathfrak G^{ij}_n (\vec k). 
\end{align}

For a non-interacting system, the integrand in Eq. \eqref{key-result} simplifies, returning $\Delta_{nm}^2 (\vec k)$ in the denominator, which cancels with $\Delta_{nm}^2 (\vec k)$ in numerator from Eq. \eqref{Qmn}. 
We thus obtain 
\begin{align}
f^{[\text{i}]}_Q (\vec q) 	 &  =  4  q^2   \sum_{ \substack { n  \in \text{filled} \\  m \ne n } }   \sum_{\vec k} \mathcal G^{xx}_{nm} (\vec k) ,
\end{align}
where 
 superscript $[\text{i}]$ means the leading (first) order contribution in $q^2$. Here  $\mathcal G^{ij}_{nm} (\vec k) = \text{Re} \mathfrak G^{ij}_{nm} (\vec k)$ is the multiband quantum metric. 
Using bound on components of quantum metric from Berry curvature $ \mathcal G^{xx}_n + \mathcal G^{yy}_n \ge |\mathcal F^{xy}_n| $,  and the definition of the Chern number of the $n$th band
\begin{align}
\sum_{\vec k} \mathcal F^{xy }_n (\vec k) = \int \frac{d^2 \vec k}{(2 \pi)^2}	 \mathcal F^{xy }_n (\vec k)  = \frac{C_n}{2 \pi}, 
\end{align}
for an occupied band with Chern number $C$ we hence receive contribution
\begin{align}
\overline{ f^{[\text{i}]}_Q }  \ge  q^2  |C|/  \pi, 
\label{leading}
\end{align}
where the overline signifies directional averaging.

Thus, the QFI at finite wavelength can be enhanced by employing high-Chern number platforms ($|C| \gg 1$), such as twisted moiré van der Waals heterostructures.

{ \textbf{Subleading-$q$ quantum Fisher information}.}  
We now turn to the $q^4$ contribution. To compute Eqs \eqref{polarization} and \eqref{geoqq}, we consider the perturbative expansion of the Bloch states in powers of $\vec{q}$:
\begin{align}
| u_{n, \vec k+ \vec q} \rangle 
= 
| u_{n, \vec k} \rangle 
+ q_{i} | \partial_{\alpha}  u_{n, \vec k} \rangle 
+ 
\frac{1}{2}  q_{i}  q_{j} |  \partial_{i} \partial_{j}
u_{n, \vec k} \rangle 
\nonumber 
\\
- \frac{1}{2} q_{i}  q_{j}
\langle u_{n , \vec k} | \partial_{i} \partial_{j}  u_{n , \vec k} \rangle 
 | u_{n, \vec k} \rangle  + \mathcal O (q ^3). 
\end{align}
The first  term in the second line ensures  normalization of Bloch states up to the third order in $q$, $\langle u_{n, \vec k+ \vec q} | u_{n, \vec k+ \vec q} \rangle  = 1 + \mathcal O (q ^3)$.  The linear term can be transformed as  
\begin{align}
|  \partial_{j}  u_{n, \vec k} \rangle  =   \sum_{p}  
| u_{p, \vec k} \rangle  
\langle u_{p, \vec k} |  
  \partial_{j}  u_{n, \vec k} \rangle 
= - i 
\sum_{p}  | u_{p, \vec k} \rangle   \mathcal A^{j}_{pn } 
\label{liner-expans}
\end{align}
where we used the resolution of identity, and the definition of the \textit{Wilczek-Zee connection}
\cite{Wilczek1984}
\begin{align}
 \mathcal A^{j}_{nm } 
 \equiv i \langle u_{n \vec k}  
|  \partial_{j}  u_{m \vec k} \rangle .
\end{align}
 Notably, the Wilczek-Zee 
 connection satisfies $\left[ \mathcal A_{nm}^j \right]^* = \mathcal A^{j}_{mn}$, a property that will prove useful in subsequent calculations. 
For the second-order terms, one thus obtains
\begin{align}
 |   \partial_{i} \partial_{j}  u_{n, \vec k} \rangle   
= - \sum_{p, p'}  
  | u_{p, \vec k} \rangle   \left [  \mathcal A^{i}_{p p' }   \mathcal A^{j}_{p' n } + i 
 \partial_{i }  \mathcal A^{j}_{pn } \delta_{p,p'}
 \right]
\end{align}
Using these  expressions, one obtains the full form of the tensor \label{geo-q} (see Supplementary Material).
\begin{widetext}
\begin{align}
\mathcal Q^{(\alpha)}_{nm} (\vec k, \vec q) & 
= |\langle u_{n, \vec k- \overline \alpha \vec q} | \mathcal V^x_{\vec k}| u_{m, \vec k+ \alpha \vec q} \rangle |^2
  =  \mathcal V^x_{nm} \mathcal V^{x*}_{nm} 
\\ 
& + 
\sum_{p, p'; s, s'}'
\Phi^*_{n; p, p'} (- \overline \alpha \vec q)  
\mathcal V^x_{ps} 
\Phi_{m; s, s'}  (\alpha \vec q)  \mathcal V^{x*}_{nm}
+ 
\sum_{p, p'; s, s'}' \mathcal V^{x}_{nm} 
\Phi^{}_{n; p, p'} (- \overline \alpha \vec q)  
\mathcal V^{x*}_{ps} 
\Phi^{*}_{m; s, s'}  (\alpha \vec q) 
\\
& + \sum_{p_1, p_1'; s_1, s_1 '}'  \sum_{p_2, p_2'; s_2, s_2'}'
\Phi^*_{n; p_1, p_1'} (- \overline \alpha \vec q)  
\mathcal V^x_{p_1 s_1} 
\Phi_{m; s_1, s_1'}  (\alpha \vec q) 
\Phi^{}_{n; p_2, p_2 '} (- \overline \alpha \vec q)  
\mathcal V^{x*}_{p_2 s_2} 
\Phi^{*}_{m; s_2, s_2'}  (\alpha \vec q)
\end{align}
\end{widetext}
Note that this expression is real by construction, and its imaginary part therefore vanishes.  Here  ${\mathcal V}^{x}_{nm}$ is a matrix element of operator $\mathcal V_{\vec k}^{x}$ in band basis ($n,m$), and quantity $\Phi_{m; s, s'}  (\vec k) $ is explicitly written below \eqref{Fqss}. 
The prime on the summation indicates the exclusion of terms with $p'=p=n$ and $s=s'=m$. One can verify that the linear-in-$q$ terms cancel due to gauge invariance. To isolate the $q^2$ contributions, we use the expansion:
\begin{widetext}
\begin{align}
\Phi_{m; s,s'} (\alpha  q) =   \delta_{s m}   \delta_{s' m}
+  \alpha q  \left[  e^{i 3 \pi /2} \delta_{s, s'}  \mathcal A^x_{s m }   \right] + \frac{1}{2} e^{i \pi} \alpha^2 q^2  (1- \delta_{sm})  \left [  \mathcal A^x_{s s' }   \mathcal A^x_{s' m } + i 
 \partial_{x }  \mathcal A^x_{s m } \delta_{ss'}
 \right]  + \mathcal O (q^3). 
 \label{Fqss}
\end{align}
\end{widetext}
The full multiband expressions, though conceptually straightforward, are technically lengthy (see Supplementary) and tend to obscure the essential structure. For clarity and brevity, we illustrate the main result using a two-band model, where the relevant terms are given by:
\begin{align}
\sum_{p,s} ' 
    \mathcal A^x_{n p} \mathcal V^x_{p s}     \mathcal A^x_{s m} \mathcal V^x_{mn}
    + 
    \mathcal V^x_{n m}  \mathcal A^x_{p n} \mathcal V^x_{s p}\, \mathcal A^x_{m s}
\end{align}
Furthermore, we can simplify calculations by using expression for the matrix elements of "anomalous velocity" $\mathcal V^x_{n m} $ in topological insulators \cite{Kruchkov2023}
\begin{align} 
\mathcal V^x_{n m}  = i \Delta_{nm} (\vec k) \mathcal A^x_{nm} (\vec k) 
\end{align}
For the two-band system, this yields 
\begin{align}
\int d \alpha \mathcal Q^{(\alpha)}_{nm} (\vec k, \vec q) &  = | \mathcal V^x_{nm}|^2 +b  \Delta_{nm}^2 
\mathcal  A^x_{nm} \mathcal  A^x_{mn} 
\mathcal  A^x_{nm} \mathcal  A^x_{mn}   + \mathcal O (q^4) 
\nonumber 
\\
& =  \Delta_{nm}^2 \mathfrak G^{xx}_{mn} +  b \Delta_{nm}^2 q^2 
[ \mathfrak G^{xx}_{mn}  ]^2   + \mathcal O (q^4) 
\end{align}
here  $b = -1/3$ comes upon $\alpha$-averaging from integral $\int _{0}^{1} d \alpha (1- \alpha ) \alpha  = 1/6$. 
Quantum geometric tensors enter naturally at this order. Accordingly, the second-order correction to the QFI is given by
\begin{align}
  f^{[\text{ii}]}_Q (\vec q)   	 &  = - \frac {4 q^4} {3}   \sum_{ \substack { n  \in \text{filled} \\  m \ne n } }     [ \mathcal G^{xx}_{nm} (\vec k) ]^2 . 
\end{align}

 { \textbf{Bounds on $q^4$ QFI}.} 
We further show the the $q^4$ term   is   bounded by Chern number squared. For simplicity, consider a two-band Chern insulator at hald filling.   The Chern number can be now considered as a "sum rule" for Berry curvature of the filled band $n$ over the Brillouin zone
\begin{align}
\int_{\text{BZ} } d^2 \vec k \, \mathcal F^{xy}_n (\vec k) = 2 \pi C.  	
\end{align}
 Application of Cauchy-Shwartz inequality therefore leads to a useful inequality 
 \begin{align}
	\int_{\text{BZ} } d^2 \vec k  \, |\mathcal F^{xy}_n (\vec k)|^2 \ge  C^2_n ,
\end{align}
where the equality is satisfied only for the Berry-flat systems.  Therefore, in discrete sum notation one get 
\begin{align}
\sum_{\vec k} \, | \mathcal F^{xy}_n (\vec k)|^2  \ge  |C_n|^2 /(2 \pi)^2
\end{align}
We further consider systems with $\sum_{\vec k} \mathcal G^{xx}_n (\vec k)  = \sum_{\vec k} \mathcal G^{yy}_n (\vec k) $. This leads, by virtue of inequalities on quantum metric $\mathcal G^{xx}_n (\vec k)  +  \mathcal G^{yy}_n (\vec k)  \ge |\mathcal F^{xy}_n (\vec k)|$. Therefore, for a filled band with Chern number $C_n = C$,  we obtain the lower bound  
\begin{align}
\sum_{\vec k}  | \mathcal G^{xx}_n (\vec k)|^2  \ge  |C|^2 /(4 \pi)^2. 
\end{align}
The subleading-order QFI is thus bounded by 
\begin{align}
| \overline{ f^{[\text{ii}]}_Q (\vec q) } | 	   \ge    \frac{ 1}{3}   \frac{q^4 C^2}{(2 \pi)^2}
\label{subleading}
\end{align}
Note that this inequality saturate in topological flat bands.

\begin{figure}[t]
\includegraphics[width = 0.9\columnwidth]{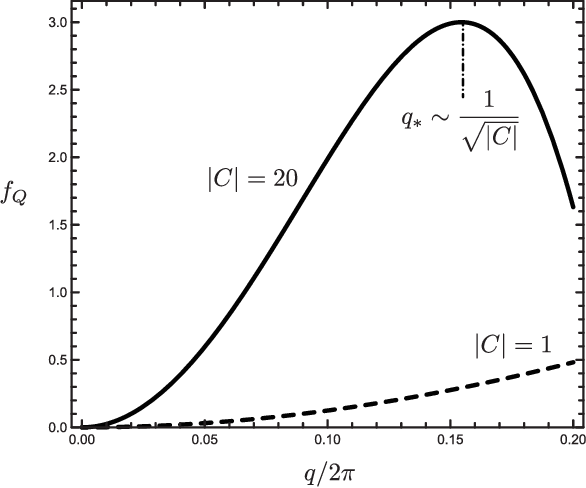}
\caption{A topological system can be engineered to enhance the QFI and control the location of its maximum at $q_* \sim 1/\sqrt{C}$. This tunability enables alignment with the profile of $V_q$ (or vice versa), maximizing operation at the quantum speed limits \eqref{speed limit}.  }
\end{figure}

\textbf{Topological control of quantum speed limits}. 
To enhance the quantum Fisher information, we propose engineering van der Waals multilayers with optimized quantum geometry. Among available strategies, increasing the number of layers in a topological heterostructure is both effective and experimentally feasible.
quite generically, the maximal Chern number in such heterostructure scales linear with the number of layers, 
$C \propto N $ \cite{Trescher2012}. Generically, this construction does not compromise band flatness, enabling the realization of nearly flat Chern bands with arbitrarily large topological index; twist angle can provide additional degree of control ~\cite{Wang2022}. Therefore, layer-by-layer stacking, when guided by quantum geometric principles, such as in twisted TMDs, offers a modular and scalable route to boosting topological response and quantum information capacity.
Crucially, the inequalities for the QFI \eqref{leading}, \eqref{subleading}  define a characteristic momentum scale  $q_* \sim \frac{2 \pi} {a \sqrt {C}} $. This suggests that heterostructures with large Chern numbers $|C| \gg 1$ can sustain high QFI at comparatively small momenta $q\sim q_* \ll \frac{2 \pi}{a}$, smaller than the size of the Brillouin zone (Fig. 1). Since Eq. \eqref{Bures} favors low-$q$ limits, this observation might be helpful for practical control over quantum speed limits.  We can  estimate \textit{bounds} on quantum speed limits from Eq. \eqref{Bures} as follows. Consider external drive in the form of $V^{\text{ext}}_{\vec q} \sim 1/q$ (e.g. such as from charged STM/AFM tip). For such a setup,  \eqref{Bures} leads to the quantum speed limit \cite{Taddei2013, Braunstein1994} of order 

\begin{align}
  \frac{ds}{dt} = \frac{1}{2}  \sqrt{\sum_{\vec{q}} f_Q(\vec{q})\, |V_{\vec{q}}^{\text{ext}}|^2 } \sim \sqrt{|C|}. 
  \label{speed limit}
  \end{align}
where we have used \eqref{leading} as a bound for QFI.  
A similar scaling with $\sqrt{|C|}$
would define characteristic rate of quantum Zeno effect \cite{Smerzi2012, Schafer2014}.

{ \textbf{Discussion}}.   
In this work, we uncover a fundamental link between quantum Fisher information (QFI) and topological invariants.   We find that even a dispersionless topological phase admits momentum-resolved QFI and the position of its local  maxima can be shifted to low momenta by system design. This further provide benefits of adjusting the external probe $V_q$ in the way that maximizes quantum speed limits for metrology. Quantum systems evolve at a finite speed that is constrained by quantum-geometric structure of the underlying quantum state. For dispersionless topological insulators, we find that the quantum speed limits scale as $\sqrt {C}$.  Systems with large Chern numbers, $|C|$$\gg$$1$, exhibit a pronounced amplification in QFI and quantum speed limits, pointing to a powerful strategy of boost the metrological precision. Twisted  multilayer van der Waals heterostructures with engineered flat bands of nontrivial quantum geometry and high Chern number are promising platforms.

\

\textit{Acknowledgments.} We thank Shinsei Ryu, Duncan Haldane, and Barry Bradlyn for very useful discussions. This work was supported by the Branco Weiss Society in Science, ETH Zurich, and Swiss National Science Foundation, Grants No. 20QU-1\_225225, CRSK-2\_221180 and Program No. IZSEZ0\_223932. 

\bibliography{Refs}

\onecolumngrid

\newpage

\section{Supplemental Material for "Topological control of quantum speed limits" }

\section{1. Relation to static structure factor}

\

\noindent
Here we shall show that integral 
\begin{align}
 f_Q(q) =  -  \int \limits_{-\infty}^{+\infty} d \omega \chi ''(\vec q, \omega) \tanh
  \left[\frac{\beta \omega}{2}
  \right]
  \label{not-Sq}
\end{align}
is NOT a static structure factor, though it looks structurally similar.

\

\noindent 
Consider the static structure factor, defined as

\begin{align}
 \  S(\vec q) = \int \limits_{-\infty}^{\infty} d \omega \, S (\vec q , \omega)  = \int \limits_{-\infty}^{\infty} d \omega \, S_{\text{sym}} (\vec q , \omega) = 2 \int \limits_{0}^{\infty} d \omega \, S_{\text{sym}} (\vec q , \omega) 
\end{align}
where $S_{\text{sym}} (\vec q , \omega) $ is the symmetric part of the dynamical structure factor, $ (\vec q , \omega)$, 
\begin{align}
S_{\text{sym}} (\vec q , \omega) 
=
\frac{S (\vec q , \omega)  + S (\vec q , - \omega)}{2}. 
\end{align}
Note that $S (\vec q , \omega) = e^{\beta \omega} S (\vec q , \omega)$, thus 
\begin{align}
S_{\text{sym}} (\vec q , \omega) 
=
S (\vec q , \omega) \frac{ 1  + e^{- \beta \omega}}{2}. 
\end{align}
Furtheron, there is a natural relation between $S_{\text{sym}} (\vec q , \omega)$ and (imaginary part of) susceptibilityA through fluctuation-dissipation theorem. Indeed, from the definition of $\chi(\vec q, t)$, we can find (we use the sign convention from the definition from Pines, Nosier, chpt. 2.7, formula (2.165-2.167)], 
\begin{align}
\chi '' (\vec q, \omega) = - \pi \left[
S (\vec q, \omega) - S (\vec q, - \omega)
\right] = - \pi S (\vec q, \omega) [1 - e^{-\beta \omega}] 
\end{align}
therefore, we have relationship between the symmetric part of $S(\vec q, \omega)$ and $\chi (\vec q, \omega)$ is given by
\begin{align}
  S_{\text{sym}} (\vec q , \omega) 
=
S (\vec q , \omega) \frac{ 1  + e^{- \beta \omega}}{2}= - \frac{1}{2 \pi}
\chi'' (\vec q, \omega)  \frac{ 1  + e^{- \beta \omega}}{ 1  - e^{- \beta \omega}} = - \frac{1}{2 \pi}
\chi'' (\vec q, \omega) \frac{ e^{ \beta \omega/2}  + e^{- \beta \omega/2}}{ e^{ \beta \omega/2}   - e^{- \beta \omega/2}} 
= - \frac{1}{2 \pi}
\chi'' (\vec q, \omega) \coth\left[ \frac{\beta \omega }{2}
\right]
\end{align}

Combining (2) with (4) 

\begin{align}
   S(\vec q) =  2 \int \limits_{0}^{\infty} d \omega \, S_{\text{sym}} (\vec q , \omega)  = - \frac{1}{\pi}  \int \limits_{0}^{\infty} d \omega \,
\chi'' (\vec q, \omega) \coth\left[ \frac{\beta \omega }{2}
\right]
\end{align}
Therefore, Eq. \eqref{not-Sq} is NOT a static structure factor, though they are structurally similar. In particular, the structure of the integrand (with hyp. cotangent) indicates that static structure factor prioritizes low-frequency, thermal fluctuation. In contrast, integrand in Eq. (1) prioritizes quantum-coherent fluctuations (higher frequency/shorter times).

\section{2. $q^4$ corrections to QFI}

\noindent 
Consider second-order perturbation theory

\begin{align}
| u_{n, \vec k+ \vec q} \rangle 
= 
| u_{n, \vec k} \rangle 
+ q^{\alpha}  \partial_{\alpha} | u_{n, \vec k} \rangle 
+ 
\frac{1}{2}  q^{\alpha}  q^{\beta} \partial_{\alpha} \partial_{\beta}
| u_{n, \vec k} \rangle  - \frac{1}{2} q^{\alpha}  q^{\beta}
\langle u_{n , \vec k} | \partial_{\alpha} \partial_{\beta}  u_{n , \vec k} \rangle 
 | u_{n, \vec k} \rangle  + \mathcal O (q ^3). 
\end{align}
The last term ensures normalization $\langle u_{n, \vec k+ \vec q}| u_{n, \vec k+ \vec q} \rangle  = 1 + \mathcal O (q ^3)$. 

The linear term can be transformed as

\begin{align}
|  \partial_{\alpha}  u_{n, \vec k} \rangle  =  \underbrace{ \sum_{p}  
| u_{p, \vec k} \rangle  
\langle u_{p, \vec k} | }_{1} 
  \partial_{\alpha}  u_{n, \vec k} \rangle 
= - i 
\sum_{p}  | u_{p, \vec k} \rangle   \mathcal A^{\alpha}_{pn } 
\end{align}
where we used the resolution of identity, and the definition of Wiczek-Zee connection 
\begin{align}
 \mathcal A^{\alpha}_{pn } 
 \equiv i \langle u_{p, \vec k}  
|  \partial_{\alpha}  u_{n, \vec k} \rangle 
\end{align}
Note that $\mathcal A_{pn}^* = \mathcal A_{np}$.  Analogously, we have

\begin{align}
\langle \partial_{\alpha}  u_{n, \vec k}  |  =   \sum_{p}  \langle \partial_{\alpha}  u_{n, \vec k}    
| u_{p, \vec k} \rangle  
\langle u_{p, \vec k} | 
= + i 
\sum_{p}  \mathcal A _{np} \langle  u_{p, \vec k} |  
\end{align}

For the second-order derivatives, we have

\begin{align}
\partial_{\alpha} |  \partial_{\beta}  u_{n, \vec k} \rangle   
= 
- i 
\partial_{\alpha } \sum_{p}  | u_{p, \vec k} \rangle   \mathcal A^{\beta}_{pn }  = 
- i 
 \sum_{p}    | \partial_{\alpha } u_{p, \vec k} \rangle    \mathcal A^{\beta}_{pn }
- i 
 \sum_{p}   | u_{p, \vec k}  \rangle  \partial_{\alpha }  \mathcal A^{\beta}_{pn }
 = -  
 \sum_{p, p'}  
  | u_{p', \vec k} \rangle   \mathcal A^{\alpha}_{p' p }   \mathcal A^{\beta}_{pn } - i 
 \sum_{p}   | u_{p, \vec k}  \rangle  \partial_{\alpha }  \mathcal A^{\beta}_{pn }
\\
 = -  
 \sum_{p, p'}  
  | u_{p, \vec k} \rangle   \mathcal A^{\alpha}_{p p' }   \mathcal A^{\beta}_{p' n } - i 
 \sum_{p}   | u_{p, \vec k}  \rangle  \partial_{\alpha }  \mathcal A^{\beta}_{pn }
= - \sum_{p, p'}  
  | u_{p, \vec k} \rangle   \left [  \mathcal A^{\alpha}_{p p' }   \mathcal A^{\beta}_{p' n } + i 
 \partial_{\alpha }  \mathcal A^{\beta}_{pn } \delta_{p,p'}
 \right]
\end{align}

Intemediate calculation:

\begin{align}
\langle u_{n , \vec k} | \partial_{\alpha} \partial_{\beta}  u_{n , \vec k} \rangle  = - \sum_{p, p'}  
 \langle u_{n , \vec k} | u_{p, \vec k} \rangle   \left [  \mathcal A^{\alpha}_{p p' }   \mathcal A^{\beta}_{p' n } + i 
 \partial_{\alpha }  \mathcal A^{\beta}_{pn } \delta_{p,p'}
 \right]
 =  - \sum_{ p'}  
   \left [  \mathcal A^{\alpha}_{n p' }   \mathcal A^{\beta}_{p' n } + i 
 \partial_{\alpha }  \mathcal A^{\beta}_{n n } \delta_{n,p'}
 \right]
\end{align}

\begin{align}
\langle u_{n , \vec k} | \partial_{\alpha} \partial_{\beta}  u_{n , \vec k} \rangle  |  u_{n , \vec k} \rangle  
 =  - \sum_{ p'}  
   \left [  \mathcal A^{\alpha}_{n p' }   \mathcal A^{\beta}_{p' n } + i 
 \partial_{\alpha }  \mathcal A^{\beta}_{n n } \delta_{n,p'}
 \right]  | u_{n , \vec k} \rangle  
\end{align}

\begin{align}
| u_{n, \vec k+ \vec q} \rangle 
= 
| u_{n, \vec k } \rangle 
+ q^{\alpha} e^{i 3 \pi /2}
\sum_{p}  | u_{p, \vec k} \rangle   \mathcal A^{\alpha}_{pn }  
+
\frac{1}{2} e^{i \pi} q^{\alpha}  q^{\beta} 
\sum_{p \ne n, p'}  
  | u_{p, \vec k} \rangle   \left [  \mathcal A^{\alpha}_{p p' }   \mathcal A^{\beta}_{p' n } + i 
 \partial_{\alpha }  \mathcal A^{\beta}_{pn } \delta_{p,p'}
 \right]
\end{align}

Now let's rewrite this in the form of 

\begin{align}
| u_{n, \vec k+ \vec q} \rangle 
= 
| u_{n, \vec k} \rangle 
+ q^{\alpha} e^{i 3 \pi /2}
\sum_{p, p'}  \delta_{p, p'} | u_{p', \vec k} \rangle   \mathcal A^{\alpha}_{p'n }  
+
\frac{1}{2} e^{i \pi} q^{\alpha}  q^{\beta} 
\sum_{p , p'}  
  | u_{p, \vec k} \rangle   \left [  \mathcal A^{\alpha}_{p p' }   \mathcal A^{\beta}_{p' n } + i 
 \partial_{\alpha }  \mathcal A^{\beta}_{pn } \delta_{p,p'}
 \right] (1- \delta_{p,n})
\end{align}

\begin{align}
| u_{n, \vec k+ \vec q} \rangle 
= 
| u_{n, \vec k} \rangle 
+
\sum_{p, p'}    | u_{p, \vec k} \rangle  \left[ q^{\alpha} e^{i 3 \pi /2} \delta_{p, p'}  \mathcal A^{\alpha}_{p n }  \right]
+
\sum_{p , p'}  
  | u_{p, \vec k} \rangle   \left [  \mathcal A^{\alpha}_{p p' }   \mathcal A^{\beta}_{p' n } + i 
 \partial_{\alpha }  \mathcal A^{\beta}_{pn } \delta_{p,p'}
 \right]  \frac{1}{2} e^{i \pi} q^{\alpha}  q^{\beta}  (1- \delta_{p,n})
\end{align}

\begin{align}
| u_{n, \vec k+ \vec q} \rangle 
= 
| u_{n, \vec k} \rangle 
+
\sum_{p, p'}    | u_{p, \vec k} \rangle  \left \{ q^{\alpha} e^{i 3 \pi /2} \delta_{p, p'}  \mathcal A^{\alpha}_{p n }  
+
 \frac{1}{2} e^{i \pi} q^{\alpha}  q^{\beta}  (1- \delta_{p,n})  \left [  \mathcal A^{\alpha}_{p p' }   \mathcal A^{\beta}_{p' n } + i 
 \partial_{\alpha }  \mathcal A^{\beta}_{pn } \delta_{p,p'}
 \right]  
 \right\}
\end{align}

This expression is normalized up to $\mathcal O (q^3)$. 

\begin{align}
f_{n; p,p'} (q) =   \left \{ q^{\alpha} e^{i 3 \pi /2} \delta_{p, p'}  \mathcal A^{\alpha}_{p n }  
+
 \frac{1}{2} e^{i \pi} q^{\alpha}  q^{\beta}  (1- \delta_{p,n})  \left [  \mathcal A^{\alpha}_{p p' }   \mathcal A^{\beta}_{p' n } + i 
 \partial_{\alpha }  \mathcal A^{\beta}_{pn } \delta_{p,p'}
 \right]  
 \right\}
\end{align}

\begin{align}
| u_{n, \vec k+ \vec q} \rangle 
= 
| u_{n, \vec k+ \vec q} \rangle 
+
\sum_{p, p'}    | u_{p, \vec k} \rangle f_{n; p, p'} (\vec q) 
\end{align}

Then we have

\begin{align}
\langle u_{n, \vec k- \overline \alpha \vec q} | \mathcal V_x| u_{m, \vec k+ \alpha \vec q} \rangle 
= 
\left( \langle  u_{n, \vec k }  | 
+
\sum_{p, p'}  f^*_{n; p, p'} (- \overline \alpha \vec q)   \langle  u_{p, \vec k} |  
\right) 
\mathcal V_x
\left( | u_{m, \vec k } \rangle 
+
\sum_{s, s'}    | u_{s, \vec k} \rangle f_{m; s, s'} ( \alpha \vec q) 
\right) 
\\
= \mathcal V_{nm}
+ \sum_{p, p'}  f^*_{n; p, p'} (- \overline \alpha \vec q) 
\mathcal V_{pm}
+\sum_{s, s'}   \mathcal V_{ns}  f_{m; s, s'} ( \alpha \vec q) 
+ \sum_{p, p'} \sum_{s, s'}  f^*_{n; p, p'} (- \overline \alpha \vec q) 
\mathcal V_{ps}
 f_{m; s, s'} ( \alpha \vec q) 
\end{align}

\begin{align}
\langle u_{n, \vec k- \overline \alpha \vec q} | \mathcal V^x_{\vec k}| u_{m, \vec k+ \alpha \vec q} \rangle 
& =  \mathcal V_{nm} \nonumber
\\
& + \sum_{p, p'} \sum_{s, s'}  f^*_{n; p, p'} (- \overline \alpha \vec q) 
\mathcal V_{ps} \delta_{s m} \delta_{s' m}
+\sum_{p, p'} m_{s, s'}  \delta_{p n} \delta_{p' n}  \mathcal V_{ps}  f_{m; s, s'} ( \alpha \vec q) 
+ \sum_{p, p'} \sum_{s, s'}  f^*_{n; p, p'} (- \overline \alpha \vec q) 
\mathcal V_{ps}
 f_{m; s, s'} ( \alpha \vec q) 
\end{align}

(all operators $\mathcal V$, $\mathcal A$ assume superscript $x$, which are dropped for notational clarity).

\begin{align}
\langle u_{n, \vec k- \overline \alpha \vec q} | \mathcal V_x| u_{m, \vec k+ \alpha \vec q} \rangle 
& =  \mathcal V_{nm}  + \sum_{p, p'} \sum_{s, s'}  f^*_{n; p, p'} (- \overline \alpha \vec q) 
\mathcal V_{ps} \delta_{s m} \delta_{s' m}
+   \delta_{p n} \delta_{p' n}  \mathcal V_{ps}  f_{m; s, s'} ( \alpha \vec q) 
+  f^*_{n; p, p'} (- \overline \alpha \vec q) 
\mathcal V_{ps}
 f_{m; s, s'} ( \alpha \vec q) 
\end{align}

Let's now write in one sum 

\begin{align}
\langle u_{n, \vec k- \overline \alpha \vec q} | \mathcal V_x| u_{m, \vec k+ \alpha \vec q} \rangle 
& =  \mathcal V_{nm}  + 
\sum_{p, p'} 
\sum_{s, s'} 
\left[ 
f^*_{n; p, p'} (- \overline \alpha \vec q)  
+ \delta_{p n} \delta_{p' n}
\right] 
\mathcal V_{ps} 
\left[
f_{m; s, s'} ( \alpha \vec q) + \delta_{s m}   \delta_{s' m}   
\right]  - \delta_{p n} \delta_{p' n}   \delta_{s m}   \delta_{s' m}
\end{align}

or 

\begin{align}
\langle u_{n, \vec k- \overline \alpha \vec q} | \mathcal V_x| u_{m, \vec k+ \alpha \vec q} \rangle 
& =  \mathcal V_{nm}  + 
\sum_{p, p'; s, s'}'
\left[ 
f^*_{n; p, p'} (- \overline \alpha \vec q)  
+ \delta_{p n} \delta_{p' n}
\right] 
\mathcal V_{ps} 
\left[
f_{m; s, s'} ( \alpha \vec q) + \delta_{s m}   \delta_{s' m}   
\right] 
\end{align}

where prime means we are excluding delta-product term with simultaneously $p=p'=n$ and $s=s'=m$ . 

Let's define 

\begin{align} 
F_{m; s, s'} (\alpha \vec q)  = f_{m; s, s'} ( \alpha \vec q) + \delta_{s m}   \delta_{s' m}
\end{align}

\begin{align}
\langle u_{n, \vec k- \overline \alpha \vec q} | \mathcal V_x| u_{m, \vec k+ \alpha \vec q} \rangle 
& =  \mathcal V_{nm}  + 
\sum_{p, p'; s, s'}'
F^*_{n; p, p'} (- \overline \alpha \vec q)  
\mathcal V_{ps} 
F_{m; s, s'}  (\alpha \vec q) 
\end{align}

Then the conjugated quantity is 

\begin{align}
\langle u_{n, \vec k- \overline \alpha \vec q} | \mathcal V_x| u_{m, \vec k+ \alpha \vec q} \rangle ^* 
& =  \mathcal V^{*}_{nm}  + 
\sum_{p, p'; s, s'}'
F^{}_{n; p, p'} (- \overline \alpha \vec q)  
\mathcal V^{*}_{ps} 
F^{*}_{m; s, s'}  (\alpha \vec q) 
\end{align}

Hence the quantity we are looking for is

\begin{align}
|\langle u_{n, \vec k- \overline \alpha \vec q} | \mathcal V_x| u_{m, \vec k+ \alpha \vec q} \rangle |^2
&  =  \mathcal V_{nm} \mathcal V^{*}_{nm} 
\\ 
& + 
\sum_{p, p'; s, s'}'
F^*_{n; p, p'} (- \overline \alpha \vec q)  
\mathcal V_{ps} 
F_{m; s, s'}  (\alpha \vec q)  \mathcal V^{*}_{nm}
+ 
\sum_{p, p'; s, s'}' \mathcal V^{}_{nm} 
F^{}_{n; p, p'} (- \overline \alpha \vec q)  
\mathcal V^{*}_{ps} 
F^{*}_{m; s, s'}  (\alpha \vec q) 
\\
& + \sum_{p_1, p_1'; s_1, s_1 '}'  \sum_{p_2, p_2'; s_2, s_2'}'
F^*_{n; p_1, p_1'} (- \overline \alpha \vec q)  
\mathcal V_{p_1 s_1} 
F_{m; s_1, s_1'}  (\alpha \vec q) 
F^{}_{n; p_2, p_2 '} (- \overline \alpha \vec q)  
\mathcal V^{*}_{p_2 s_2} 
F^{*}_{m; s_2, s_2'}  (\alpha \vec q)
\end{align}

Note, that this expression is real by construction, hence it's imaginary part is zero.

\

\textbf{Linear terms}

($\alpha= \beta = x$, index dropped)

\begin{align}
F_{n; p,p'} (q) =   \delta_{p n}   \delta_{p' n}+  q  \left[  e^{i 3 \pi /2} \delta_{p, p'}  \mathcal A_{p n }   \right] 
\end{align}

\begin{align}
F_{m; s,s'} (q) =   \delta_{s m}   \delta_{s' m}
+  q  \left[  e^{i 3 \pi /2} \delta_{s, s'}  \mathcal A_{s m }   \right] 
\end{align}

there will be six terms.  One can check that linear terms vanish term by term upon $\alpha$ integration. Indeed, (27) has e.g. term

\begin{align}
\sum_{p, p'; s, s'}'
F^*_{n; p, p'} (- \overline \alpha \vec q)  
\mathcal V_{ps} 
F_{m; s, s'}  (\alpha \vec q)  \mathcal V^{*}_{nm}
 \rightarrow
\underbrace{ \sum_{p, p'; s, s'}
 \delta_{p n}   \delta_{p' n}  
\mathcal V_{ps} 
\alpha q    e^{i 3 \pi /2} \delta_{s, s'}  \mathcal A_{s m }    \mathcal V^{*}_{nm}
}_{-i \alpha q     \sum_{ s}
\mathcal V_{ns} 
 \mathcal A_{s m }    \mathcal V_{mn}}
\\
 + \underbrace{  \sum_{p, p'; s, s'}
(- \overline \alpha  q )    e^{-i 3 \pi /2} \delta_{p, p'}  \mathcal A^*_{p n }     
\mathcal V_{ps} 
\delta_{s m}   \delta_{s' m}   \mathcal V^{*}_{nm}
}
_
{
(- i \overline \alpha  q )  \sum_{p}
  \mathcal A_{np }     
\mathcal V_{pm} 
  \mathcal V_{mn}
}
\end{align}

Upon averaging over $\alpha$, we thus obtain

\begin{align}
\int_{0}^{1} d\alpha \sum_{p, p';\, s, s'}
 F^*_{n; p, p'}(-\overline{\alpha} \vec{q}) \, 
\mathcal{V}_{p s} \, 
F_{m; s, s'}(\alpha \vec{q}) \, 
\mathcal{V}^*_{n m}
\rightarrow\ 
&\, - \frac{i q}{2}\left [ \sum_{s} \mathcal{V}_{n s} \, \mathcal{A}_{s m} \, 
+ \sum_{s} \mathcal{A}_{n s} \, \mathcal{V}_{s m} 
\right] \mathcal{V}_{m n}
\\
&=  - \frac{ q}{2}\left [ \sum_{s} \Delta_{sn} \mathcal{A}_{n s} \, \mathcal{A}_{s m} \, 
+  \Delta_{ms} \mathcal{A}_{n s} \, \mathcal{A}_{s m} 
\right] \mathcal{V}_{m n}
\\
& =  - \frac{ q}{2}\left [ \sum_{s} \underbrace{
(\Delta_{sn} + \Delta_{ms} ) 
}_{\Delta_{mn}}
\mathcal{A}_{n s} \, \mathcal{A}_{s m} 
\right] \mathcal{V}_{m n}
\\
&  =+ \frac{i q \Delta_{mn}^2}{2}\left [ \sum_{s} 
\mathcal{A}_{n s} \, \mathcal{A}_{s m} 
\right] \mathcal{A}_{m n}
\end{align}

The second term in (27) gives 

\begin{align}
\sum_{p, p'; s, s'} \mathcal V^{}_{nm} 
F^{}_{n; p, p'} (- \overline \alpha \vec q)  
\mathcal V^{*}_{ps} 
F^{*}_{m; s, s'}  (\alpha \vec q) 
 \rightarrow
\underbrace{ \sum_{p, p'; s, s'} \mathcal V^{}_{nm} 
 \delta_{p n}   \delta_{p' n}  
\mathcal V^{*}_{ps} 
  \left[ \alpha q  e^{-i 3 \pi /2} \delta_{s, s'}  \mathcal A*_{s m }   \right]
}_{
i \alpha q  \mathcal V^{}_{nm} 
\sum_{s} \mathcal V_{s n} 
 \mathcal A_{ms }  
}
\\
 + \underbrace{ \sum_{p, p'; s, s'} \mathcal V^{}_{nm} 
(- \overline \alpha q )   \left[  e^{i 3 \pi /2} \delta_{p, p'}  \mathcal A_{p n }   \right] 
\mathcal V^{*}_{ps} 
 \delta_{s m}   \delta_{s' m}
}
_
{
i \overline \alpha q  \mathcal V^{}_{nm}   \sum_{p} 
  \mathcal A_{p n }  
\mathcal V_{mp}
}
\end{align}

\begin{align}
\int \limits_{0}^{1 } d \alpha \sum_{p, p'; s, s'} \mathcal V^{}_{nm} 
F^{}_{n; p, p'} (- \overline \alpha \vec q)  
\mathcal V^{*}_{ps} 
F^{*}_{m; s, s'}  (\alpha \vec q) 
 \rightarrow
\frac{i q}{2} \mathcal V^{}_{nm} 
\sum_{s}' \mathcal V_{s n} 
 \mathcal A_{ms }  
+  \sum_{s} 
  \mathcal A_{s n }  
\mathcal V_{ms}
\\
=
\frac{ q}{2} \mathcal V^{}_{nm} 
\sum_{s}  \Delta_{ns } \mathcal A_{s n} 
 \mathcal A_{ms }  
+  
  \mathcal A_{s n }  
\Delta_{s m} \mathcal A_{ms}
\\
=
\frac{ q}{2} \mathcal V^{}_{nm} 
\sum_{s}  \Delta_{mn } \mathcal A_{s n} 
 \mathcal A_{ms }  
 \\
=
-\frac{i q \Delta_{mn }^2}  {2} \mathcal A^{}_{nm} 
\left[ \sum_{s}  \mathcal A_{s n} 
 \mathcal A_{ms }   \right] =  -\frac{i q \Delta_{mn }^2}  {2} \mathcal A^{*}_{mn} 
\left[ \sum_{s}  \mathcal A_{s n} 
 \mathcal A_{ms }   \right] 
\end{align}
Note that this expression does not contribute in the linear order in q due to exclusion of summation in $n,m$. 
We can further use identities  $A_{nm}^2 = \sum_{p}A_{np} A_{pn}$, hence linear terms  vanish.

\

\textbf{Quadratic terms }

\begin{align}
f_{n; p,p'} (q) =   \left \{ q^{\alpha} e^{i 3 \pi /2} \delta_{p, p'}  \mathcal A^{\alpha}_{p n }  
+
 \frac{1}{2} e^{i \pi} q^{\alpha}  q^{\beta}  (1- \delta_{p,n})  \left [  \mathcal A^{\alpha}_{p p' }   \mathcal A^{\beta}_{p' n } + i 
 \partial_{\alpha }  \mathcal A^{\beta}_{pn } \delta_{p,p'}
 \right]  
 \right\}
\end{align}

(for calculation below, $\alpha= \beta = x$, index dropped)

\begin{align}
F_{n; p,p'} ( - \overline \alpha \vec q) =   \delta_{p n}   \delta_{p' n} - \overline \alpha q  \left[  e^{i 3 \pi /2} \delta_{p, p'}  \mathcal A_{p n }   \right] 
+ \frac{1}{2} e^{i \pi} {\overline \alpha} ^2 q^2  (1- \delta_{p,n})  \left [  \mathcal A_{p p' }   \mathcal A_{p' n } + i 
 \partial_{x}  \mathcal A_{pn } \delta_{p,p'}
 \right]  
\end{align}

\begin{align}
F^{*}_{n; p,p'} ( - \overline \alpha \vec q) =   \delta_{p n}   \delta_{p' n} - \overline \alpha q  \left[  e^{-i 3 \pi /2} \delta_{p, p'}  \mathcal A_{np }   \right] 
+ \frac{1}{2} e^{- i \pi} {\overline \alpha} ^2 q^2  (1- \delta_{p,n})  \left [  \mathcal A_{p' p }   \mathcal A_{n p' } - i 
 \partial_{x}  \mathcal A_{np } \delta_{p,p'}
 \right]  
\end{align}

\begin{align}
F_{m; s,s'} (\alpha \vec q) =   \delta_{s m}   \delta_{s' m}
+  \alpha q  \left[  e^{i 3 \pi /2} \delta_{s, s'}  \mathcal A_{s m }   \right] + \frac{1}{2} e^{i \pi} \alpha^2 q^2  (1- \delta_{s,m})  \left [  \mathcal A_{s s' }   \mathcal A_{s' m } + i 
 \partial_{x }  \mathcal A_{s m } \delta_{s,s'}
 \right]  
\end{align}

\begin{align}
F^{*}_{m; s, s'}  (\alpha \vec q)  =   \delta_{s m}   \delta_{s' m}
+  \alpha q  \left[  e^{-i 3 \pi /2} \delta_{s, s'}  \mathcal A_{ m s }   \right] + \frac{1}{2} e^{-i \pi}  \alpha^2 q^2  (1- \delta_{s,m})  \left [  \mathcal A_{s' s }   \mathcal A_{m s'} - i 
 \partial_{x}  \mathcal A_{m s } \delta_{s,s'}
 \right]  
\end{align}

\begin{align}
\sum_{p, p'; s, s'}'
F^*_{n; p, p'} (- \overline \alpha \vec q)  
\mathcal V_{ps} 
F_{m; s, s'}  (\alpha \vec q)  \mathcal V_{mn}
+ 
\sum_{p, p'; s, s'}' \mathcal V^{}_{nm} 
F_{n; p, p'} (- \overline \alpha \vec q)  
\mathcal V_{sp} 
F^{*}_{m; s, s'}  (\alpha \vec q) 
\\
 + \sum_{p_1, p_1'; s_1, s_1 '}'  \sum_{p_2, p_2'; s_2, s_2'}'
F^*_{n; p_1, p_1'} (- \overline \alpha \vec q)  
\mathcal V_{p_1 s_1} 
F_{m; s_1, s_1'}  (\alpha \vec q) 
F_{n; p_2, p_2 '} (- \overline \alpha \vec q)  
\mathcal \mathcal V_{s_2 p_2} 
F^{*}_{m; s_2, s_2'}  (\alpha \vec q)
\end{align}

There will be several terms 

\begin{align}
\mathcal S 
=
q^2 \bar\alpha^2 \mathcal S^{(\bar\alpha^2)}
+ 
q^2 \alpha^2  \mathcal S^{(\alpha^2)}
 + 
q^2  \alpha \bar\alpha \mathcal S^{(\alpha \bar\alpha)}.
\end{align}

or, upon averaging, just

\begin{align}
\mathcal S 
=
\frac{q^2}{4}  \mathcal S^{(\bar\alpha^2)}
+ 
\frac{q^2}{4}  \mathcal S^{(\alpha^2)}
 + 
\frac{q^2}{6} \mathcal S^{(\alpha \bar \alpha)}.
\end{align}

\begin{align}
\mathcal  S^{(\bar\alpha^2)} & 
= 
\frac12 e^{-i\pi}
\sum_{p ,p'} '
\Bigl[
   \mathcal  A_{p'p} \mathcal A_{n p'} - i \partial_x \mathcal A_{n p} \delta_{p,p'}
\Bigr]\,
\mathcal V_{p m} \mathcal V_{ mn }
+
\frac12 
e^{i\pi}
\sum_{p ,p'} ' V_{n m}  
\Bigl[
    \mathcal A_{p p'} \mathcal A_{p' n} + i \partial_x \mathcal A_{p n} \delta_{p,p'}
\Bigr] 
\mathcal V _{m p} 
+ \sum_{p ,p'} '
\mathcal A_{n p}\, \mathcal V_{p m} \mathcal A_{p' n} \mathcal V_{m p'},
\end{align}

\begin{align}
\mathcal S^{(\alpha^2)} &
= 
\frac12 
e^{i\pi}
\sum_{s ,s'}' \mathcal V_{n s} 
\Bigl[
    \mathcal A_{s s'} \mathcal A_{s' m} + i \partial_{x} \mathcal A_{s m} \delta_{s,s'}
\Bigr] 
\mathcal V_{m n }
+
\frac12\,e^{-i\pi}
\sum_{s,s'}'
\mathcal V_{n m} 
\Bigl[
    \mathcal A_{s' s} \mathcal A_{m s'} - i \partial_x \mathcal A_{m s} \delta_{s,s'}
\Bigr] 
\mathcal V_{s n}
\\
&
+\sum_{s,s'}'
\mathcal V_{n s}\, \mathcal A_{s m} \mathcal V_{s' n} \mathcal A_{ m s'},
\end{align}

\begin{align}
\mathcal S^{\alpha \bar\alpha}
&
=
- \sum_{p,s} ' 
\Bigl(
    \mathcal A_{n p} \mathcal V_{p s} \mathcal A_{s m} \mathcal V_{mn}
    + 
    \mathcal V_{n m} \mathcal A_{p n} \mathcal V_{s p} \mathcal A_{m s}
\Bigr)
+ \sum_{p,s} '
\Bigl(
    \mathcal A_{n p} \mathcal V_{p m} \mathcal V_{s n} \mathcal A_{ms}
     + 
   \mathcal  V_{n s} \mathcal A_{s m} \mathcal A_{p n} \mathcal V_{m p}
\Bigr).
\end{align}

For the two-band system, (95-99) gives:

\begin{align}
\mathcal S^{\alpha \bar\alpha}  = -  \frac{q^2}{3} \Delta_{nm}^2 
\mathcal A_{nm} \mathcal A_{mn} 
\mathcal A_{nm} \mathcal A_{mn}   =  -  \frac{q^2}{3} \Delta_{nm}^2 
[ \mathcal G_{mn}^{xx} ]^2
\end{align}

\

\end{document}